\documentclass[prb,showpacs,superscriptaddress,twocolumn]{revtex4}
\usepackage{graphicx,amsmath,amsfonts,amssymb}

\usepackage{epstopdf}
\usepackage[utf8]{inputenc}
\usepackage{color}  
\usepackage{array}

\RequirePackage[
   hyperindex,colorlinks,bookmarksnumbered,
   plainpages=true,pdfstartview=FitH]{hyperref}
\hypersetup{linkcolor=blue,urlcolor=blue,citecolor=blue}
\usepackage{hyperref}
\usepackage{ulem}

\renewcommand{\emph}[1]{\textit{#1}}
\newcommand{\TK}{T_{\rm K}}
\newcommand{\TKa}{T_{{\rm K}\alpha}}
\newcommand{\TKr}[1]{T_{\rm K}^{(#1)}}
\newcommand{\TKrn}{\TKr{n}}
\newcommand{\Hloc}{\hat{H}_\mathrm{loc} }
\newcommand{\JH}{J_{\rm H}}
\newcommand{\rhom}{\rho_{\rm m}}
\newcommand{\rhomNRG}{\rho_{\rm m}^{\rm NRG}}
\newcommand{\rhoph}{\rho_{\rm ph}}
\newcommand{\gammam}{\gamma_{\rm m}}
\newcommand{\SUn}{{\rm SU}(n)}
\newcommand{\Etrunc}{E_{\rm trunc}}
\newcommand{\Eq}[1]{Eq.~(\ref{#1})}

\newcommand{\pprI}{paper~I}

\begin{document}

\title{
  Iron impurities in gold and silver: Comparison of transport
  measurements to numerical renormalization group calculations
  exploiting non-Abelian symmetries
}

\author{M. Hanl}
\author{A. Weichselbaum}
\affiliation{Physics Department, Arnold Sommerfeld Center for Theoretical Physics and Center for NanoScience, Ludwig-Maximilians-Universit\"at M\"unchen, 80333 M\"unchen, Germany}
\author{T. A. Costi}
\affiliation{Peter Gr\"unberg Institut and Institute for Advanced Simulation, 
Research Centre J\"ulich,  52425 J\"ulich, Germany}
\author{F. Mallet}
\affiliation{Institut N{\'e}el-CNRS and Universit{\'e} Joseph Fourier, 38042 Grenoble, France}
\author{L. Saminadayar}
\affiliation{Institut N{\'e}el-CNRS and Universit{\'e} Joseph Fourier, 38042 Grenoble, France}
\affiliation{Institut Universitaire de France, 103 boulevard Saint-Michel, 75005 Paris, France}
\author{C. B\"auerle}
\affiliation{Institut N{\'e}el-CNRS and Universit{\'e} Joseph Fourier, 38042 Grenoble, France}
\author{J. von Delft}
\affiliation{Physics Department, Arnold Sommerfeld Center for Theoretical Physics and Center for NanoScience, Ludwig-Maximilians-Universit\"at M\"unchen, 80333 M\"unchen, Germany}
\date{May 14, 2013}

\begin{abstract}
  We consider iron impurities in the noble metals gold and silver and
  compare experimental data for the resistivity and decoherence rate
  to numerical renormalization group results.  By exploiting
  non-Abelian symmetries we show improved numerical data for both
  quantities as compared to previous calculations [Costi \emph{et
  al.}, Phys.~Rev.~Lett.~\textbf{102}, 056802 (2009)], using
  the discarded weight as criterion to reliably judge the
  quality of convergence of the numerical data. In addition we also
  carry out finite-temperature calculations for the magnetoresistivity
  of fully screened Kondo models with $S = \frac{1}{2}$, 1 and
  $\frac{3}{2}$, and compare the results with available measurements
  for iron in silver, finding excellent agreement between theory and
  experiment for the spin-$\frac{3}{2}$ three-channel Kondo
  model. This lends additional support to the conclusion of Costi
  \textit{et al.}~that the latter model provides a good effective
  description of the Kondo physics of iron impurities in gold and
  silver.

\end{abstract}
\maketitle

The magnetic alloys for which the Kondo effect was first
observed, in the 1930s, were iron impurities in gold and silver
\cite{deHaas1934,deHaas1936}.  They showed an anomalous rise in the
resistivity with decreasing temperature, which Kondo explained in 1964 as being
due to an antiferromagnetic exchange coupling between the localized
magnetic impurity spins and the spins of the delocalized conduction
electrons \cite{Kondo1964}. For his work, Kondo used a
spin-$\frac{1}{2}$, one-band model, which undoubtedly captures the
essential physics correctly in a qualitative way. 

However, detailed comparisons between theory and experiment have
since shown that this model does not yield a \textit{quantitatively}
correct description of the Kondo physics of dilute Fe impurities in
Au or Ag. Such a description must meet the challenge of
quantitatively reproducing, using the Kondo temperature $\TK$ as
only fitting parameter, several independent sets of experimental
measurements: the contributions by magnetic impurities (indicated by
a subscript m) to the temperature- and field-dependence of the
resistivity, $\rhom(T, B)$, and to the temperature-dependence of the
decoherence rate, $\gammam(T)$, extracted from weak
(anti)localization measurements. The spin-$\frac{1}{2}$, 1-band
Kondo model does not meet this challenge: when comparing its
predictions, obtained by the numerical renormalization group
(NRG)\cite{Wilson1975,Krishnamurthy1980,Bulla2008}, to transport
measurements on dilute Fe impurities in Ag wires, different Kondo
scales were required for fitting the resistivity and decoherence
rates \cite{Mallet2006,Alzoubi2006}.

In a recent publication (Ref.~\onlinecite{Costi2009}, involving most
of the present authors, henceforth referred to as paper I), it was
argued that the proper effective low-energy Kondo model for Fe in Au
or Ag is, in fact, a fully screened, spin-$\frac{3}{2}$ three-channel
Kondo model. \pprI\ arrived at this conclusion by the following chain
of arguments.  Previous transport
experiments\cite{Mallet2006,Alzoubi2006} had indicated that these
systems are described by a fully screened Kondo
model\cite{Nozieres1980,Andrei1984,Tsvelik1985,Affleck1990,Affleck1992},
i.e.\ a Kondo model in which the local spin, $S$, is related to the
number of conduction bands, $n$, by $S=n/2$. As mentioned above, the
choice $n=1$ had already been ruled out in earlier work
\cite{Mallet2006,Alzoubi2006}. Density-functional theory
calculations for Fe in Au and Ag, presented in paper~I, showed that
in these host metals Fe preferentially acts as a substitutional
defect with cubic symmetry, leading to a substantial crystal field
splitting ($\ge 0.15$~eV) between a higher-lying $e_g$ doublet and a
lower-lying $t_{2g}$ triplet.  Moreover, the local spin moment was
predicted to be 3 Bohr magnetons, with an almost fully quenched orbital
angular momentum. This suggested a fully-screened Kondo model with
$n=3$ as the most likely candidate, while leaving some scope for the
possibility of $n=2$ (but none for $n=4$ or 5).  To discriminate
between the options $n=2$ and 3, $\rhom(T,0)$ and $\gammam(T)$ were
then calculated using NRG, for $n=1$ (as reference), 2 and 3. Next,
for both material systems (Fe in Au and Ag), the $\rhom(T,0)$ curves
were fitted to experimental data to obtain a Kondo temperature,
$\TKrn$, for each of the three models.  Finally, using these
$\TKrn$-values, the $\gammam(T)$ curves, which constituted
parameter-free predictions of the decoherence rate, were compared to
corresponding measurements, with the conclusion that the choice $n=3$
worked distinctly better than $n=2$.

The goal of the present paper is two-fold: 
First, we describe technical details of the 
numerical calculations performed in paper~I that
could not be presented  there for lack of space.
Second and more important, we extend the analysis of paper~I
to the case of finite magnetic fields.  Indeed, though experimental
data for $\rhom(T,B \neq 0)$ had been available for Fe in Ag even at
the time of writing of paper~I, it had not been possible then to
compare them to theoretical predictions for $n=3$. The reason is
that multichannel calculations present an enormous challenge for the
NRG, as the numerical complexity grows exponentially with the number
of channels. In \pprI{} only Abelian symmetries (charge conservation
in each channel and total spin $S_z$) were exploited. For the
purposes of \pprI, this turned out to be sufficient, but for the
aforementioned three-channel Kondo model the calculations were
numerically extremely costly, and even at $B=0$ just barely within
the limits of feasibility. When the present authors
attempted, in subsequent work (unpublished), to treat the more
general case of a finite magnetic field using the same approach, the
latter turned out to be inadequate, plagued by numerical convergence
issues. Therefore, further progress required enhancing the numerical
efficiency by exploiting non-Abelian symmetries.

Now, the effective fully screened symmetric three-channel Kondo model
mentioned above has several \emph{non}-Abelian symmetries, including,
in particular, an SU(3) channel symmetry. This implies that the
eigenspectrum of the Hamiltonian can be organized into degenerate
symmetry multiplets, and great gains in numerical efficiency can be
made by exploiting this multiplet structure at every step of the NRG
procedure.  We took this observation as incentive to implement
non-Abelian symmetries in our code on a completely generic footing for
tensor networks such as the NRG.\cite{Weichselbaum2012} Although the
exploitation of symmetries, Abelian as well as non-Abelian, together
with their respective strong gain in numerical efficiency is well
known in the literature, the treatment of non-Abelian symmetries in
NRG has been largely restricted to the symmetry of SU(2)
\cite{Wilson1975,Krishnamurthy1980,Toth2008,Toth2008a}.  The
non-Abelian symmetry SU(2), however, is simpler than the general case,
since for $n \geq 3$ the $\SUn$ representation theory involves
complications due to the presence of inner and outer multiplicities.
A generic numerical framework for treating arbitrary non-Abelian
symmetries thus had been missing, and became available only very
recently.\cite{Weichselbaum2012,Alex2011,Moca2012}

More specifically, the model Hamiltonians studied here possess SU(2)
particle-hole symmetry, $\SUn$ channel symmetry, and SU(2) spin
symmetry for $B=0$ or Abelian $S_z$ symmetry for $B \neq 0$.  By
exploiting the non-Abelian symmetries, we were able to
drastically reduce the computational effort and generate fully
converged numerical data, even for the highly challenging case of
three channels. With a significantly more powerful NRG at our hands
then, the following analysis serves two purposes. First, we present a
thorough reanalysis of \pprI{} with improved NRG data. In
particular, we give a detailed discussion of NRG truncation and
convergence issues, which are under much better control with the new
non-Abelian scheme. The new numerical results show
discernible quantitative differences w.\,r.\,t.\ \pprI, leading to
changes in the deduced Kondo temperatures that are quite substantial
for $n=3$ (the relative change in $\TK$ is 31\% for Fe in Au and
53\% for Fe in Ag). Second, we present a detailed analysis of the
new numerical magnetoresistivity data and compare these to experimental
results for Fe in Ag.  The results of both analyses fully confirm the
main conclusion of \pprI{}: the effective microscopic model for dilute
iron impurities in the noble metals gold and silver is given by a
fully screened three-channel Kondo model.

The remainder of this paper is organized as follows:
Sec.~\ref{sec:model} describes the model, Sec.~\ref{sec:NRGdetails}
describes NRG-related details. Sec.~\ref{sec:comparison} provides a
comparison of experimental and numerical magnetoresistance
data, followed by a summary in Sec.~\ref{sec:conclusions}.

\section{Model}
\label{sec:model}
In paper~I we found it numerically convenient for our NRG
calculations to start not from a pure Kondo model but from an
effective Anderson-type model, because it is then possible to obtain
an improved spectral function by using the so-called ``self-energy
trick'' \cite{Bulla1998}, which involves calculating the
impurity-level self-energy. It has recently been
shown\cite{Mitchell2011} that a similar strategy can 
be used for Kondo-type models, but this fact was brought
to our attention only after completion of the present
study \cite{referee}.

We here adhere to the strategy of paper I and adopt
the following Anderson-type model,
\begin{eqnarray}
  \hat{H} & = &     \sum_{\alpha=1}^n \sum_{k\sigma} 
    \left( 
       t (\hat{d}^{\dag}_{\alpha\sigma}
       \hat{c}_{k\alpha\sigma}^{\phantom\dagger}
     + \mathrm{H.c.})
     + \varepsilon_k \hat{c}^{\dag}_{k\alpha\sigma}
       \hat{c}_{k\alpha\sigma}^{\phantom\dagger}
    \right) \nonumber
\\
& &  - \JH^{(n)}\hat{\vec S}^2_{\rm imp} + g \mu_{\rm B} B \hat{S}^z_{\rm imp} \; , 
\label{eq:Kondo_AM}
\end{eqnarray}
which reduces to a Kondo-type model at low
energies\cite{Muehlschlegel1968,Nishikawa2012}. The index $\alpha$ labels $n$
degenerate local levels as well as $n$ independent channels of
conduction electrons, each forming a flat band of half-bandwidth $D=1$
with constant density of states $\nu_0 = 1/{2D}$ per spin and
channel. (In the remainder of the paper, all energies are specified
in units of half-bandwidth, unless indicated otherwise.)
${\hat d}_{\alpha\sigma}$ is the annihilation operator of an
impurity electron with spin $\sigma$ in level $\alpha$, whereas ${\hat
  c}_{k\alpha\sigma}$ annihilates a reservoir-electron in channel
$\alpha$ with wave number $k$ and energy $\varepsilon_k$. Levels and
channels are tunnel-coupled diagonally in spin and channel indices,
resulting in a width $\Gamma=\pi\nu_0t^2$ for each level, $t$ being
the hopping matrix element between impurity and reservoir.
The third term in $\hat H$ describes a Hund-type exchange interaction
  with $\JH^{(n)}>0$, added to favor a local spin of $S=n/2$, where
$\hat{\vec S}_{\rm imp}=\sum_{\alpha=1}^n \hat{\vec S}_{\alpha}$ is the total
impurity spin operator, $\hat{\vec S}_{\alpha}=\frac{1}{2}
\sum_{\sigma\sigma'}\hat{d}_{\alpha\sigma'}^{\dag} \vec \tau_{\sigma'\sigma}
\hat{d}_{\alpha\sigma}$ is the spin operator for an electron in level
$\alpha$, and $\vec \tau=(\tau_x,\tau_y,\tau_z)$ are Pauli spin
matrices. The last term in $\hat H$ describes the effect of an
applied local magnetic field, with $g = 2$.
To ensure particle-hole
symmetry (which renders the numerics more efficient) we take
$\varepsilon_{\alpha\sigma}=0$ for the local level positions and do
not include any further charging energy.  

The energies of the free orbital (FO) states are given by
roughly $\JH^{(n)} S(S+1)$ and the energy difference between two
FO-states that differ by spin $\frac{1}{2}$ is therefore given by
$\Delta E^{(n)} \approx \JH^{(n)} [(S(S+1)-(S-\frac{1}{2})(S+\frac{1}{2})]
=\JH^{(n)}(S+\frac{1}{4})$. To focus on the local moment regime of
the Anderson model, we choose $\JH^{(n)}$ such that $\Delta E^{(n)}$
is significantly larger than $\Gamma$ and $g\mu_{\rm B} B$, ensuring a
well-defined local spin of $S=n/2$, and an average total occupancy
of the local level of $\sum_{\alpha \sigma} \langle
\hat{d}^\dagger_{\alpha \sigma} \hat{d}^{}_{\alpha \sigma}\rangle = n$.
Moreover, the ratios $\JH^{(n)}/\Gamma$ are chosen such that the
resulting Kondo temperatures have comparable magnitudes.

In paper~I, we had implemented this strategy using the same $\Delta
E^{(n)}$ for all three $n$-values, with $\Gamma = 0.01$ and
$\JH^{(1)} = 0.053$, $\JH^{(2)} = 0.032$, $\JH^{(3)} = 0.023$. We
have since realized that much better NRG convergence properties can
be obtained by choosing much larger values of $\JH^{(n)}$, to ensure
that the energy differences of the FO states truly lie well above
the bandwidth ($\Delta E^{(n)} \gtrsim 100$).  This is the numerical
counterpart to a Schrieffer-Wolff transformation
\cite{Schrieffer1966,Weichselbaum2012}: it shifts the numerically
most expensive, yet irrelevant, FO-regime to an energy range which
lies outside the range whose energies are finely resolved during the
NRG diagonalization, thus reducing the numerical costs needed for
treating the Anderson model to a level comparable to that of the
Kondo model.
For the numerical calculations presented here, we set the level
width to $\Gamma=25$ and choose $\JH^{(n)}$ such that the resulting
spectral functions have 
the same half-width at half maximum $(=2\times10^{-4})$
for all three cases, $n\in \{1, 2, 3\}$, thus ensuring that the
Kondo temperatures are equal. This is achieved by choosing the Hund
couplings as $\JH^{(1)}=358.9$, $\JH^{(2)}=112.8$, and $\JH^{(3)}=57.14$.

For the model in Eq.~(\ref{eq:Kondo_AM}), the resistivity and decoherence 
rate due to magnetic impurities (relevant for weak localization) can be 
calculated as follows \cite{Micklitz2006,Zarand2004}:\
\begin{align}
\label{eq:def_rhom}
&\rhom^{\rm NRG}(T,B)=\frac{\rhom^0}{2n} \int d\omega f'(\omega) \sum_{\alpha\sigma}{\rm Im}(\Gamma G^R_{\alpha\sigma}(\omega)), \\
\label{eq:def1_gammam}
&\gammam^{\rm NRG}(T)=\left[ \int d\omega [-f'(\omega)] \sqrt{\gammam(\omega,T)} \right]^2, \\
\label{eq:def2_gammam}
&\gammam(\omega,T)= -\frac{\gammam^0}{2n} \sum_{\alpha\sigma} \left[ {\rm Im}(\Gamma G^R_{\alpha\sigma}(\omega)) + \vert \Gamma G^R_{\alpha\sigma}(\omega) \vert^2 \right].
\end{align}
Here $G^R_{\alpha\sigma}(\omega)$ is the fully-interacting retarded
impurity Green's function, $f'(\omega)$ is the derivative of the Fermi
function, $\rhom(0)=\rhom^0=2\tau {\bar \rho}/\pi \hbar \nu_0$ and
$\gammam^0=2/\pi\hbar\nu_0$, where ${\bar \rho}$ is the
resistivity due to static disorder and $\tau$ the corresponding 
elastic scattering time. For \emph{real} materials with complex 
Fermi surfaces, both prefactors $\rhom^0$ and $\gamma_{\rm m}^0$ 
contain material-dependent (hence unknown) factors arising from 
integrals involving the true band structure of the conduction electrons.\\

\section{NRG details}\label{sec:NRGdetails}

\subsection{Wilson chain and spectral function} 

Within the NRG, the non-interacting bath in \Eq{eq:Kondo_AM}
is coarse grained using the dimensionless discretization
parameter $\Lambda > 1$, followed by the mapping
onto the so-called Wilson chain in terms of the fermionic Wilson
sites \cite{Wilson1975, Krishnamurthy1980, Bulla2008}
$\hat{f}_{i'\alpha\sigma}$ with $i'\in\{0,1,\ldots\}$. 
Therefore, $\hat{H} \cong \lim_{N\to\infty} \hat{H}_N$, where
\begin{subequations}
\begin{eqnarray}
  \hat{H}_N \cong \Hloc &+&
       \sum_{i'=0}^{N-1} t_{i'}
       \sum_{\alpha=1}^n \sum_{\sigma} 
       (\hat{f}^{\dag}_{i',\alpha\sigma}
        \hat{f}_{i'+1,\alpha\sigma}^{\phantom\dagger}
     + \mathrm{H.c.}) \qquad \phantom{.}
     \label{eq:Kondo_AM:Wilson}
\end{eqnarray}
with
\begin{eqnarray}
\Hloc &\equiv& \hat{H}_{\rm J} + 
    \sum_{\alpha=1}^n \sum_{\sigma} 
       \sqrt{\tfrac{2\Gamma}{\pi}}
      (\hat{d}^{\dag}_{\alpha\sigma}
       \hat{f}_{0\alpha\sigma}^{\phantom\dagger}
     + \mathrm{H.c.})
\end{eqnarray}
where
\begin{equation}
\hat{H}_{\rm J}\equiv - \JH^{(n)}\hat{\vec S}^2_{\rm imp} + g \mu_{\rm B} B \hat{S}^z_{\rm imp}\text{.}
\end{equation}

\end{subequations}
The impurity spin is coupled to a semi-infinite tight-binding
chain with the exponentially decaying couplings $t_{i'} \propto \Lambda^{-i'/2}$.
For large enough $\Lambda \gtrsim 2$, this ensures energy scale
separation, and thus justifies the iterative diagonalization of
the Hamiltonian in the representation of the Wilson chain.
\cite{Wilson1975, Krishnamurthy1980, Bulla2008}
In particular, the energies of the Hamiltonian
$\hat{H}_i$ at intermediate iterations which include all terms $i' < i$,
are rescaled in units of $\omega_i$, where
\begin{equation}
\omega_i \equiv a\Lambda^{-i/2}
\text{.}\label{eq:EScale}
\end{equation}
Here the constant $a$ is chosen such that $\lim_{i\to\infty}
t_i/\omega_i=1$ An analytic expression for $a$ in the presence of
$z$-shifts is given in Ref.~\onlinecite{Weichselbaum2011}.

To obtain the Green's function $G^R_{\alpha\sigma}(\omega)$, which
determines $\rhom^{\rm NRG}(T,B)$ and $\gammam^{\rm NRG}(T)$, we
calculate the spectral function $A_{\alpha\sigma}(\omega)=-\frac{1}{\pi}{\rm
Im}(G^R_{\alpha \sigma}(\omega))$ using its Lehmann representation:
\begin{equation}
\label{eq:def_Aomega}
A_{\alpha\sigma}(\omega)=\sum_{a,b}\frac{e^{-\beta E_a}+e^{-\beta E_b}}{Z}
\vert \langle a \vert \hat{d}_{\alpha\sigma} \vert b \rangle \vert^2
\delta (\omega-E_{ab}),
\end{equation}
where $E_{ab}=E_b-E_a$, with $E_a$, $E_b$ and $\vert a \rangle$,
$\vert b \rangle$ being the eigenenergies and many-body
eigenstates obtained by NRG in the full density matrix (FDM)-approach
\cite{Weichselbaum2007,Anders2005,Peters2006,Weichselbaum2012a}.
Note that due to the $\SUn$ symmetry of the
Hamiltonian, the spectral function $A_{\alpha\sigma}(\omega)$
does not depend on the index $\alpha$.
Thus when exploiting non-Abelian symmetries,
in practice, one calculates the channel-independent symmetrized spectral function
$A_{\sigma}(\omega)
\equiv \frac{1}{n}\sum_{\alpha=1}^n A_{\alpha\sigma}(\omega)$,
which corresponds to the normalized scalar contraction
$\hat{d}_\sigma^\dagger \cdot \hat{d}_\sigma
\equiv \sum_\alpha \hat{d}_{\alpha\sigma}^\dagger \cdot
\hat{d}_{\alpha\sigma}$
of the spinors $\hat{d}_\sigma$. \cite{Weichselbaum2012}

\label{sec:methods}
  
For the calculation of $\gammam(T)$ the knowledge of both the real and
the imaginary part of $G^R_{\alpha\sigma}(\omega) \equiv G^R_{\sigma}(\omega)$
is necessary. The real part can be determined via the Kramers-Kronig relations from
$A_{\sigma}(\omega)$ after smoothing the discrete data. $\rhom^{\rm
NRG}(T,B)$, on the other hand, requires only the imaginary part of
the Green's function. This makes the application of the Kramers-Kronig
relations and with it the broadening of the discrete data unnecessary
and $\rhom^{\rm NRG}(T,B)$ can therefore be directly calculated from
the discrete data \cite{Weichselbaum2007}, thus avoiding
possible broadening errors. Furthermore, due to particle-hole symmetry,
it is sufficient to calculate $A_{\sigma}(\omega)$ only for one spin
$\sigma$, since the spectral functions for opposite spins $\sigma$ and
${\bar \sigma}$ are symmetric with respect to each other:
$A_{\sigma}(\omega)=A_{\bar \sigma}(-\omega)$.

\subsection{Convergence criteria and discarded weight}

\begin{figure}[tb!]
\includegraphics{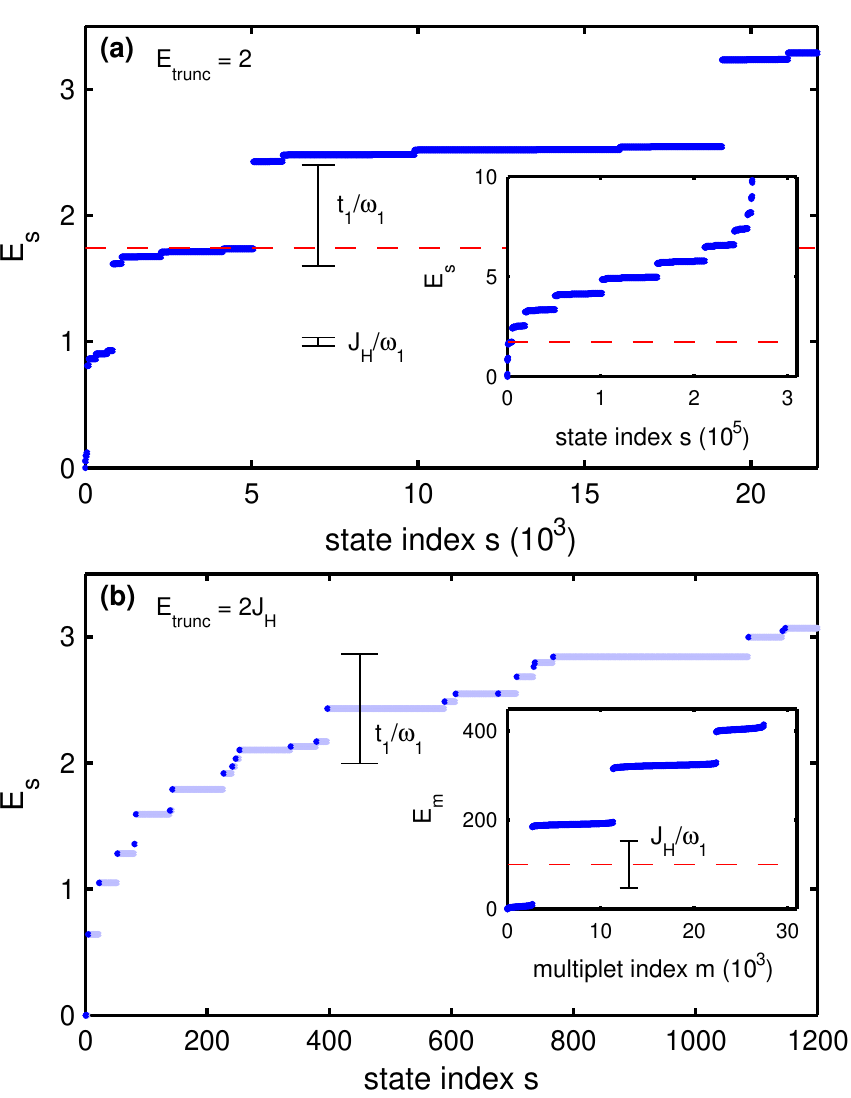}
\caption{(Color online) Eigenenergies of the $n=3$ calculations from
  (a) \pprI{} and (b) this work, for the lowest eigenstates (blue
  circles) and truncation energy (dashed red line) of NRG iteration
  $i=1$. This iteration includes the impurity and the first two Wilson
  sites $\hat{f}_0$ and $\hat{f}_1$, which by \Eq{eq:Kondo_AM:Wilson}
  corresponds to $\hat{H}_1$; it is the first iteration where
  truncation occurred. All energies $E_s$ are given in units of
  $\omega_1$ [c.f. \Eq{eq:EScale}]. In (a), each dark blue dot marks
  an eigenstate; in panel (b) each dark blue dot marks a multiplet,
  whose degeneracy is indicated by the length of the adjacent light
  blue lines.  Dashed red lines indicate the truncation energy
  $\Etrunc$. In paper~I, the number of kept states at iteration
  $i=1$ was 4840 which was 216 states short of truncating into
  the wider energy gap starting at $E_s=5056$.
  For the present paper, we chose the truncation energy to lie well
  within a wide spectral gap and kept 16 384 out of 262 144 states 
  [only a small subset of which are shown in the main
  panel of (b)]. This large number was achievable by grouping
  the kept states into 2,688 symmetry multiplets with internal degeneracy.
  The insets of (a) and (b) show, respectively, the full spectrum of states or
  multiplets at iteration $i=1$. (The fine structure seen in the main
  panel in (b) is not resolved in the inset, since the latter uses a
  much coarser energy resolution on the vertical axis.) The spectra in
  (a) and (b) have different fine structure, because the model
  parameters were chosen differently in \pprI{} and the present work,
  respectively: the former used $\JH^{(3)}=0.0229$, $\Gamma=0.01$, the
  latter $\JH^{(3)}=57.14$, $\Gamma=25$. As a result, the energy
  separation between degenerate multiplets at the truncation energy is
  different, namely $\mathcal{O}(t_1/\omega_1)$ in (a) versus
  $\mathcal{O}(\JH^{(n)}/\omega_1)$ in (b), where $t_1$ is the hopping
  matrix element between the first two sites of the Wilson chain
  [c.f. \Eq{eq:Kondo_AM:Wilson}]. The different values of $\JH$ and
  $t_1$ used in (a) and (b) are indicated by black lines in the
  plots. \vspace{-0.3in} }
\label{fig:kept-states}
\vspace{-0.2in} 
\end{figure}

As mentioned in the introduction, when using Abelian symmetries the
calculations described above are standard for $n=1$ and $n=2$, but a
real challenge for $n=3$.  
The reason is that the number of degenerate eigenstates in a
typical symmetry multiplet increases strongly with the rank of the
symmetry group.  For example, for the present model with $n=3$, the
typical degeneracy quickly becomes of order $10^2$ to $10^3$ even
for low-lying energy multiplets (this is illustrated by the presence
of long ``plateaux'' in the excitation spectra shown in Fig.~1
below). This implies that the number of kept states needs to
increase dramatically, too. Moreover a crucial prerequisite for
well-converged results is that the multiplet structure should be
respected during NRG truncation. No multiplet should be kept only
partially, i.e.\ cut in two; instead, each multiplet should be kept
or discarded as a whole.  In the present paper, cutting multiplets
is avoided by implementing non-Abelian symmetries explicitly and
keeping all multiplets below a specified truncation energy, as
described further below. In paper~I, which implemented only Abelian
symmetries, we had used the more conventional NRG truncation scheme
of specifying the total maximum number of states to typically be
kept. However, we had adjusted this number as needed to ensure that
the lowest-lying discarded states were not degenerate with the
highest-lying kept states. Moreover, the energy
of the highest kept multiplet turned out to lie just below
a wide gap in the energy spectrum [see Fig.~\ref{fig:kept-states}(a)].
In our subsequent work we have found that the presence of this wide gap 
considerably stabilizes the results; when we keep some more 
multiplets such that the highest ones lie just
above the wide gap, the results deteriorate considerably,
as judged by the criterion discussed next.

The criterion used in \pprI\ to judge the quality of convergence was
based on the Friedel sum rule\cite{Langreth1966}, which for the
present model implies that the Kondo peak of the zero-temperature
spectral function should satisfy $\pi\Gamma \cdot
A_{\alpha\sigma}(\omega=0)=1$. For \pprI\ this check was satisfied to
within 2\% for spectral functions calculated using the self-energy
trick, which we had taken as indication that the data could be
trusted. When calculated \textit{without} the self-energy trick,
though, the Kondo peak height was off by 1\%, 16\% and 32\% for
$n=1$, 2 and 3, respectively, which, in retrospect, indicates lack
of full convergence for the latter two cases.

Indeed, this became apparent \textit{a posteriori} in the
course of the present study when we reanalyzed the NRG data of
\pprI\ using a more reliable tool for checking NRG convergence that
had been developed in 2011\cite{Weichselbaum2011}, based on
monitoring the discarded weight. In essence, the discarded
weight measures the relevance of the highest-lying kept states for
obtaining an accurate description of the ground state space a few
iterations later. More concretely, it is calculated as follows:
construct a reduced density matrix for a chain of length $i$ from
the mixed density matrix of the ground state space of a chain of
length $i+i_0$ by tracing out the last $i_0$ sites (typically
$i_0 \gtrsim 4$ to ensure that all eigenvalues of the reduced density matrix are non-zero);
find the eigenvalues and eigenstates of this reduced
density matrix, say $\rho_r^{[i;i_0]}$ and $|r_{i;i_0}\rangle$,
and sort them according to their energy expectation values,
$E_r^{[i;i_0]} = \langle r_{i;i_0} | \hat H_i | r_{i;i_0}\rangle$.
The weight $\varepsilon_{5\%,i}^{\rm D}
\cong \sum^{\rm top
  \, 5\%}_r \rho_r^{[i;i_0]}$ contributed by the highest-lying 5\% of
states in this energy-sorted list then provides an estimate
for the discarded weight at iteration
$i$. It provides a quantitative measure for the importance of the
discarded states had they been included in the description of the
ground state space of iteration $i+i_0$ by keeping a larger number
of states. Repeating this analysis for different sites
$i$, the largest $ \varepsilon_{5\%,i}^D$ value  
is taken to define the ``discarded weight'' 
of the entire Wilson chain, $\varepsilon_{5\%}^{\rm D} = \max^{}_i(
\varepsilon_{5\%,i}^D)$. The entire analysis concerns the kept space only, hence
it is fast relative to the actual NRG calculation itself. Well-converged physical
quantities are obtained when the discarded weight satisfies
$\varepsilon_{5 \%}^{\rm D} \lesssim 10^{-10}$. For the NRG data
used in paper~I, the discarded weight calculated \textit{a
posteriori} turned out to be $2.8\times10^{-13}$, $2.9\times10^{-9}$
and $8.3 \times 10^{-7}$ for $n=1, 2$ and $3$, respectively. This
indicates lack of proper convergence for $n=2$, and especially for
$n=3$.
 
\subsection{Truncation scheme for non-Abelian symmetries}
For the calculations presented here,
we therefore use an improved code, which also exploits non-Abelian
symmetries \cite{Weichselbaum2012}. Here, the idea is to make use of
the fact that degenerate states can be gathered into symmetry
multiplets.  By the Wigner-Eckart theorem, matrix elements including
states from the same multiplet are then related via Clebsch Gordan
coefficients. Thus, it is sufficient to keep track not of all
individual states inside each multiplet, but only of entire
multiplets, and to store only one reduced matrix element for each
multiplet. This drastically reduces the size of the matrix which has
to be diagonalized at an NRG iteration, with corresponding reductions
in calculation times and memory requirements.

Our model possesses the following non-Abelian symmetries: ${\rm
SU}(2)$ particle-hole symmetry, ${\rm SU}(2)$ spin symmetry (in the
absence of magnetic field) and $\SUn$ channel symmetry. For many
of our calculations, we need $B \neq 0$, in which case the ${\rm
SU}(2)$ spin symmetry is reduced to an Abelian symmetry using
$S_z$. Moreover, particle-hole symmetry and channel symmetry do not
commute in general, yet their combination generates the larger
symplectic symmetry ${\rm Sp}(2n)$
(Ref.~\onlinecite{Weichselbaum2012}). This symmetry, which
encompasses both particle-hole and channel symmetry, fully exhausts
the model's symmetry; consequently no degeneracies remain between
different ${\rm Sp}(2n)$ multiplets (a typical multiplet contains
several hundreds up to several thousands of states).  For the
calculations presented in this work, using $\SUn$ [rather than ${\rm
Sp}(2n)$] turned out to be sufficient. Here we use $\SUn$ channel
symmetry together with total charge for $n \in \{2, 3\}$ and
particle-hole symmetry for $n=1$. The gain in numerical efficiency due
to these symmetries is huge. For example, for $n=3$, the largest
$\SUn$ multiplets kept in our NRG calculations already reach
dimensions of above 100.  By exploiting these symmetries,
calculation times as well as memory requirements are reduced by more
than two orders of magnitude compared to those of paper I. As a
consequence, the calculations presented here can be simply performed
within a few hours on standard workstations.

We used an NRG discretization parameter of $\Lambda=4$, and perform
$z$-averaging \cite{Yoshida1990} with $N_z=2$ (and $z \in \{ 0, 0.5
\}$) to minimize discretization artifacts \cite{Merker2013}. For
$n=3$, the computationally most challenging case, we used the
following truncation scheme.  For the diagonalization of $H_0 \equiv
\Hloc$, all states were kept. For iteration $i=1$, we used a
truncation energy [given in rescaled units of $\omega_{i=1}$, c.f.\
\Eq{eq:EScale}] of $\Etrunc =2\JH/D>7$.  Fig.~\ref{fig:kept-states}(b)
shows a subset of the corresponding kept eigenenergies and multiplet
degeneracies, while Fig.~\ref{fig:kept-states}(a) shows corresponding
information for the calculations from \pprI. The inset of
Fig.~\ref{fig:kept-states}(b) shows that \textit{all} Kondo-like
states of the Anderson model have been retained.
For iterations $i \geq 2$, we
used $\Etrunc=7$, except for $z=0.5$ at iteration $i=2$, where we used 
$\Etrunc=6$ to reduce computational costs due to the
extraordinary large density of states at that iteration; this choice of
parameters corresponds to keeping $\lesssim 10\;000$ multiplets
($\lesssim 77\;000$ states).
Using this scheme, a single NRG run for $n=3$
required about 40 GB of RAM and took on the order of 10 hours of
calculation time on an 8-core processor. The subsequent calculation of
the spectral function required a similar amount of time and 55 GB
memory.  The large number of kept multiplets then resulted in high
numerical accuracy. In particular, the spectral functions calculated
with and without using the improved self-energy, already agreed very
well with each other, which clearly demonstrates fully converged
numerical data. Having established this for a few representative
cases, we proceeded to calculate the data presented below without
using the self-energy trick.

\begin{figure}
\includegraphics{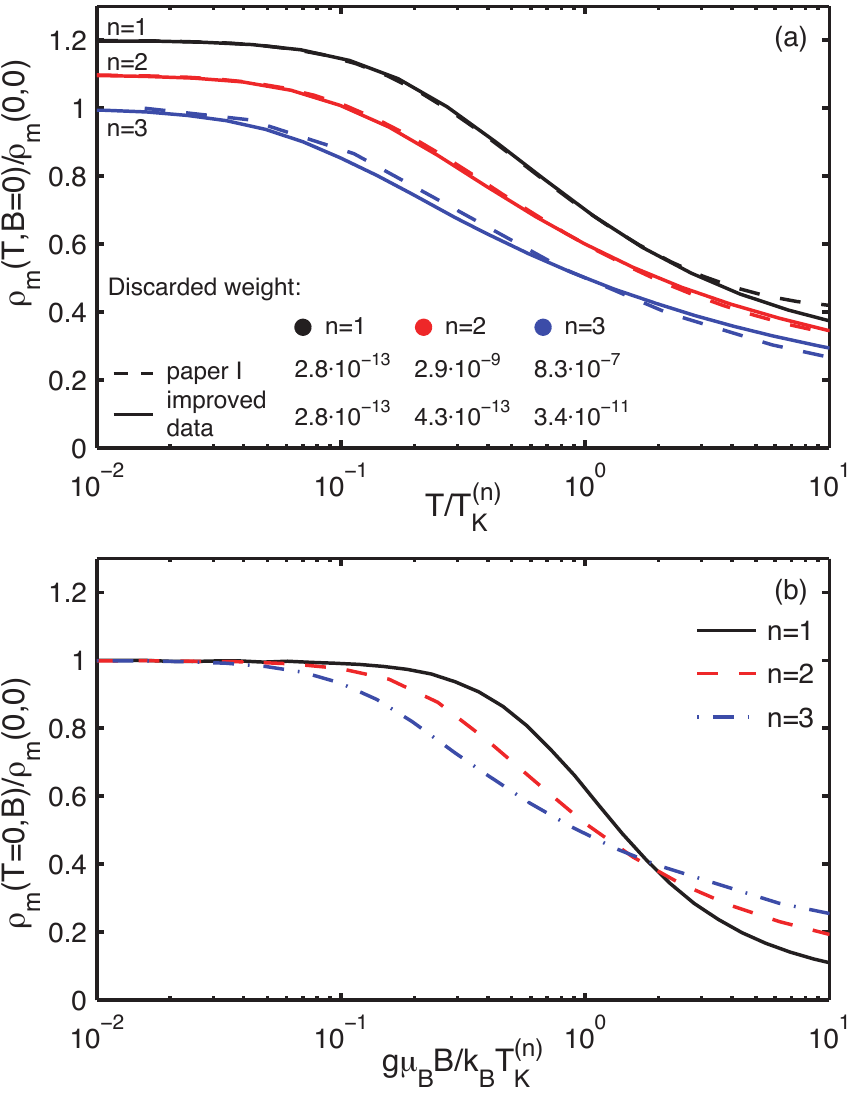}
\caption{(Color online) (a) 
Temperature dependence of the zero-field
  resistivity for $n=1$, 2 and 3, computed using both Abelian NRG with
  self-energy trick as in \pprI\ (dashed lines) and  our new
  non-Abelian NRG approach (solid lines). For clarity,
  successive curves have been vertically shifted by 0.1.
  (b) The magnetic-field dependence of the zero-temperature resistivity
  for $n=1$, 2 and 3, calculated using non-Abelian NRG.
}
\label{fig:compare-old-new}
\end{figure}

\subsection{Resistivity obtained by non-Abelian NRG}

To compare the results obtained with our new approach with those of
paper~I, Fig.~\ref{fig:compare-old-new}(a) shows the temperature
dependence of the zero-field resistivity for $n=1$, 2 and 3, computed
using both Abelian NRG with self-energy trick as in \pprI\ (dashed
lines) and using our new non-Abelian NRG approach (solid lines), which
produced truly well-converged results.  
We define the Kondo temperature
$\TKrn$ associated with a given numerical resistivity curve
$\rhomNRG(T,0)$ by the condition
\begin{eqnarray}
\label{eq:define-Tk-resistivity}
\rhomNRG(\TKrn,0)= {\textstyle \frac{1}{2}} \rhomNRG(0,0) \; .
\end{eqnarray}
Then $\rhomNRG (T,0)/\rhomNRG (0,0)$ vs.\ $T/\TK$ should be a universal
curve for given $n$. For $n=1$ and 2 the solid and dashed lines in
Fig.~\ref{fig:compare-old-new}(a) agree well (except at large
temperatures for $n=1$, where the dashed curve is affected by
free-orbital states, implying that in paper~I, $\TK^{(1)}$ had not
been chosen sufficiently small w.r.t.~the FO excitation energy). For
$n=3$, however, the shapes of the dashed and solid curves actually
differ quite noticeably. The reason for the difference is the lack
of full convergence of the Abelian NRG data.  This becomes clearly
evident by comparing the discarded weights, listed in the legend of
Fig.~\ref{fig:compare-old-new}(a), of the non-Abelian and Abelian
calculations: for $n=3$, the respective discarded weights of $3.4
\times 10^{-11}$ and $8.3 \times 10^{-7}$ indicate that the former
calculations, but not the latter, are well converged.  This comparison
thus highlights both the benefits of exploiting non-Abelian symmetries
in order to reduce convergence problems, and the importance of
checking the latter in a reliable fashion by monitoring the discarded
weight.

The fact that the resistivity curve for $n=3$ shows a more gradual decrease with increasing temperature
for the new non-Abelian results than for the old Abelian ones,
implies that fits to experiment will yield a larger Kondo
temperature for the former, as we indeed find
below.

Fig.~\ref{fig:compare-old-new}(b) shows the zero-temperature
magnetoresistivity curves for $n=1$, 2 and 3, calculated by
non-Abelian NRG.  
The curves are scaled by the same $\TKrn$
as derived from the temperature-dependent data where the latter,
by construction, cross at $T=\TKrn$
[cf. \Eq{eq:define-Tk-resistivity}]. Interestingly, 
the magnetic-field dependent curves here also
approximately cross a common point at a magnetic field
of about $g\mu_B B \sim 1.8\,k_{\rm B}\TKrn$ 
having $\rho_m(T=0,B)/\rho_m(0,0) \simeq 0.4$. 
The general trend of the curves in Fig.~\ref{fig:compare-old-new}(b) 
is similar to that seen in
Fig.~\ref{fig:compare-old-new}(a): the larger $n$, the more gradual
the decrease in resistivity with increasing temperature or field.
This indicates that the larger the local spin $S=n/2$, the larger
the energy range (in units of $\TK^{(n)}$) within which its
spin-flip-scattering effects are felt strongly by conduction
electrons. In absolute energy units, this tendency is even stronger, since
the fits to experiment performed below yield $\TK^{(1)} < \TK^{(2)}
< \TK^{(3)}$ (cf.~Table \ref{tab:fit_parameters}).  Interestingly, the $n$-dependent differences in curve
shapes are more pronounced for the field dependence than for the
temperature dependence\cite{analytics}: in
Fig.~\ref{fig:compare-old-new}(b) the decrease of the resistivity
for a given $n$ sets in at a higher energy and then is steeper than
in Fig.~\ref{fig:compare-old-new}(a).  Thus, the comparison between
experiment and theory for the magnetoresistivity performed below
constitutes a stringent test of which choice of $n$ is most
appropriate, independent of and complementary to the tests performed
in \pprI.

\section{Comparison with experiment}
\label{sec:comparison}
To identify the microscopic model which describes the system of iron
impurities in gold and silver correctly, we compare NRG calculations
for the resistivity $\rhom^{\rm NRG}(T,B)$ and the decoherence rate
$\gammam^{\rm NRG}(T)$ to experimental data, $\rhom^{\rm exp}(T,B)$
and $\gammam^{\rm exp}$. [In the following when referring to both NRG
and experiment, we omit the upper index and write $\rhom(T,B)$ and
$\gammam(T)$.] The data to be analyzed stems from a detailed
experimental study\cite{Mallet2006} performed in 2006 on
quasi-one-dimensional wires.  One AuFe-sample and two AgFe-samples
were studied, to be denoted by AuFe3, AgFe2 and AgFe3, with impurity
concentrations of $7 \pm 0.7$, $27 \pm 3$ and $67.5 \pm 7$ ppm,
respectively. These concentrations are so small
  that multi-impurity effects can be ignored. Low-field measurements
of the temperature-dependence of the resistivity, performed at
$B=0.1$~T to suppress weak localization, are available for all three
samples.  We will denote this data by $\rhom^{\rm exp}(T,0)$ [rather
than $\rhom^{\rm exp}(T,0.1{\rm T})$], and compare it to numerical
results for $\rhom^{\rm NRG}(T,0)$ computed at $B=0$, since
$1-\rhom^{\rm NRG}(T,0.1{\rm T})/\rhom^{\rm NRG}(T,0)<0.5\%$ for all three cases
$n\in\{1,2,3\}$. Moreover, experimental data is available for $\gammam^{\rm exp}(T)$
from AgFe2 and  AuFe3, and for $\rhom^{\rm exp}(T,B)$ from AgFe2.

The comparison between experiment and theory proceeds in three steps:
(i) First, we compare measured data and NRG predictions for the
resistivity at zero magnetic field $\rhom(T,B=0)$ to determine two fit
parameters, $\TKrn$ and $\delta^{(n)}$, for each of the samples and
each of the three models $n \in \{1, 2, 3\}$. After the fit parameters
have been determined, we use $\TKrn$ and $\delta^{(n)}$ to make
parameter-free predictions for (ii) the decoherence rate
$\gammam(T)$ and (iii) the temperature-dependent
magnetoresistivity $\rhom(T,B)$, and compare these to experiment for
those samples for which corresponding data is available. Here (i)
and (ii) represent a thorough reanalysis of the experimental data of
\pprI\ using our new, improved numerical data, while (iii) involves
experimental data not published previously, and new numerical
data.

\begin{table}
\begin{center}
\setlength\extrarowheight{4pt}
\begin{tabular}{ l  c  c  c  c } \hline\hline 
                                               & $n$ &               AuFe3 &                 AgFe2 &         AgFe3         \\\hline
 $\TKr{n}$                                     & 1 & \ \ $0.6 \pm 0.1$ \ \ & \ \ $2.5 \pm 0.2$ \ \ & \ \ $2.8 \pm 0.2$ \ \ \\
 $({\rm K})$                                   & 2 &     $1.0 \pm 0.1$     &     $4.3 \pm 0.3$     &     $4.7 \pm 0.3$     \\
                                               & 3 &     $1.7 \pm 0.1$     &     $7.4 \pm 0.5$     &     $8.2 \pm 0.5$     \\\hline
 $\delta^{(n)}$                                & 1 &            -0.002     &         0.003         &         0.001         \\
 $({\rm n\Omega \cdot cm/ppm})$                & 2 &            -0.045     &        -0.005         &        -0.007         \\
                                               & 3 &            -0.090     &        -0.013         &        -0.016         \\\hline
 $\Delta\rho^{\rm exp}(0,0)$                   &   &             0.211     &         0.041         &         0.041         \\
 $({\rm n\Omega \cdot cm/ppm})$                &   &                       &                       &                       \\\hline
 $\rhom^{{\rm uni},(n)}(0,0)$                  & 1 &             0.213     &         0.038         &         0.040         \\
 $({\rm n\Omega \cdot cm/ppm})$                & 2 &             0.256     &         0.046         &         0.048         \\
                                               & 3 &             0.301     &         0.054         &         0.057         \\\hline\hline
\end{tabular}
\caption{Values of parameters determined from fitting the experimental
  measurement. The values for $\TKrn$ and
  $\delta^{(n)}$ are extracted using  the
  fitting procedure whose results are shown in
  Fig.~\ref{fig:rho(T)}. $\Delta\rho^{\rm exp}(0,0)$ is the
  measured value for the resistivity at zero magnetic field and
  the lowest temperature avalaible.  For the sake of
  completeness, we also show $\rhom^{{\rm
  uni},(n)}(0,0)=\Delta\rho^{\rm exp}(0,0)-\delta^{(n)}$,
  which, according to \Eq{eq:rhoNRG}, corresponds to the unitary
  Kondo resistivity.}
\label{tab:fit_parameters}
\end{center}
\end{table}

\begin{figure}
\includegraphics{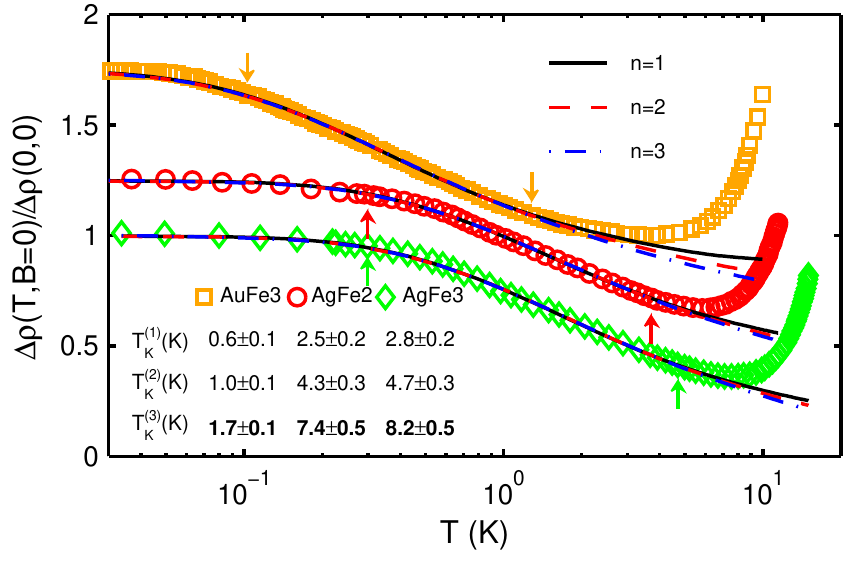}
\caption{(Color online) Similar figure as Fig.~3 of \pprI{}, but using
  substantially improved numerical data. The figure shows low-field
  experimental data for the temperature-dependence of the resistivity,
  denoted by $\Delta\rho^{\rm exp}(T,0)$ but taken in a small field of
  0.1~T to suppress weak localization (see text), and NRG calculations
  for $n \in \{ 1, 2, 3 \}$, performed at $B=0$.  The NRG curves were
  fitted to the experimental data, using $\TKrn$ and $\delta^{(n)}$ as
  fitting parameters [see Eq.~(\ref{eq:rhoNRG})] with the fitting
  range being indicated by arrows. For temperatures below the fitting
  range, the data are less reliable due to a long equilibration time,
  whereas for temperatures above the fitting range the
  phonon-contribution to $\Delta\rho^{\rm exp}(T,B=0)$ becomes
  relevant. For clarity, the curves for AgFe2 and AuFe3 have been
  shifted vertically by 0.25 and 0.75, respectively.}
\label{fig:rho(T)}
\end{figure}

\subsection{Determination of fit parameters}

The experimental resistivity data to be discussed below (shown in
Fig.~\ref{fig:rho(T)}) has several contributions of different
physical origin:
\begin{equation}
\label{eq:Delta_rho^exp}
\Delta\rho^{\rm exp}(T,B) = \rhom^{\rm exp}(T,B) + \rhoph(T) + 
\delta \, . 
\end{equation}

Here $\rhom^{\rm exp}(T,B)$ is the resistivity due to magnetic
impurities, $\rhoph(T)$ is the resistivity due to phonon scattering,
and $\delta$ is an unknown offset which does not depend on temperature
or magnetic field. There are two further contributions to the
resistivity: a classical contribution,\cite{Alzoubi2006} which
scales as $B^2$, and a contribution due to electron-electron
interactions, \cite{Altshuler1985,Akkermans2007} which scales as
$1/\sqrt{T}$. These have already been subtracted from the measured
resistivity data shown in Figs.~\ref{fig:rho(T)} and
\ref{fig:rho(T,B)} using procedures described in
Refs.~\onlinecite{Bauerle2005,Saminadayar2007}, and hence are not
displayed in \Eq{eq:Delta_rho^exp}.

For the fitting process at $B=0$, the normalized NRG data
$\rhomNRG(T,0)/\rhomNRG(0,0)$ are approximated by a fitting function
$g_n(T/\TKrn)$ constructed from higher-order polynomials, where
$g_n(0)=1$ and $\TKrn$ is fixed by scaling the temperature axis such
that $g_n(1) = \frac{1}{2}$ 
[cf.\ \Eq{eq:define-Tk-resistivity}]. 
We then fit the experimental data to the
form
\begin{equation}
\label{eq:rhoNRG}
\Delta\rho^{\rm exp}(T,0) \approx \delta^{(n)}
+[\Delta\rho^{\rm exp}(0,0) - \delta^{(n)}] g_n(T/\TKrn),
\end{equation}
using a $\chi^2$-minimization with $\TKrn$ and $\delta^{(n)}$ as fit
parameters. While a similar analysis was performed in \pprI{}, the
numerical data in the present paper are of improved quality, in
that we can report fully converged data also for the numerically
extremely challenging case of $n=3$. The newly extracted values of
$\TKrn$ for the three samples are given in Table~\ref{tab:fit_parameters}. 
For $n \in \{1, 2\}$ they are slightly different from the ones of 
\pprI{}, yet within the given error bars 
(14 \% and 0 \% for AuFe3, 9 \% and 5 \% for AgFe, respectively) due to 
the fact that we used different fitting ranges to minimize the error arising 
from the phonon-contribution for larger $T$ and because we use higher-order 
polynomials to approximate the NRG data, which may be considered more accurate 
than the analytical expression used in \pprI{}. The
difference in $\TK$ is more substantial for $n=3$ (31 \% for AuFe3 and
53 \% for AgFe) reflecting larger differences between the previous and
our new, improved NRG results for $n=3$. Experimental and fitted
NRG data are shown in Fig.~\ref{fig:rho(T)}.

\begin{figure}
\includegraphics{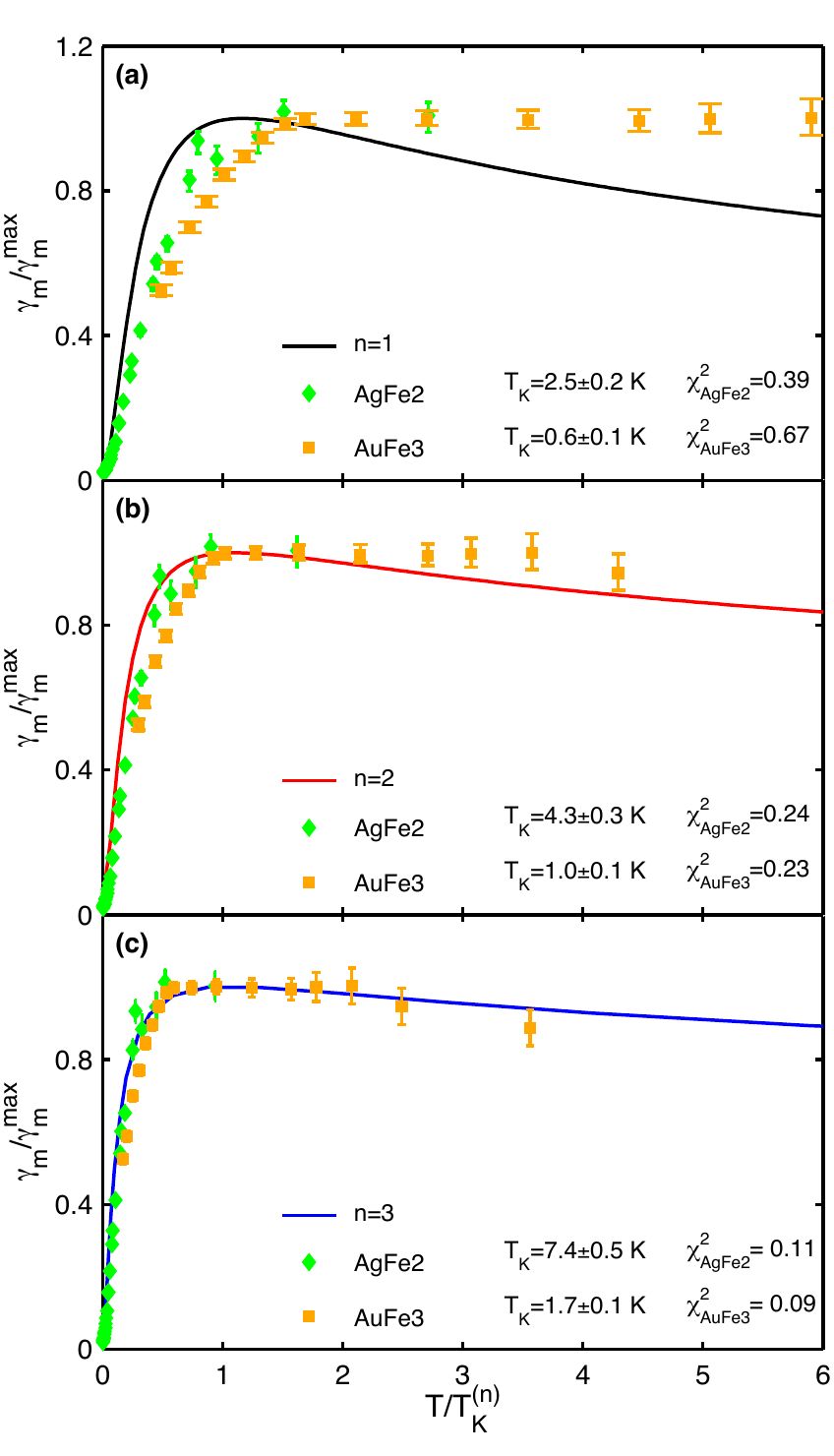}
\caption{(Color online) Similar figure as Fig.~4 of \pprI{}, but using clearly
  improved numerical data. Panels (a), (b) and (c) show the normalized
  decoherence rate $\gammam (T)/ \gammam^{\rm max}$ vs.~$T/\TKrn$ for
  $n \in \{1, 2, 3\}$, respectively. The Kondo temperatures are
  determined from the fits of $\rhom^{\rm NRG}(T,B=0)$ to the
  experimental data according to Eq.~(\ref{eq:rhoNRG}). The $\chi^2$-values 
  indicated in the legends were obtained as the sum of the least squares
  between the experimental data and the linearly interpolated
  NRG curves.}
\label{fig:gammam(T)}
\end{figure}

\subsection{Decoherence rate and magnetoresistivity}

With the $\TKrn$ for AgFe2 and AuFe3 determined above we are now in a
position to make a parameter-free theoretical prediction of the
decoherence rate.  As shown in Fig.~\ref{fig:gammam(T)} for AgFe2 and
AuFe3, the agreement is clearly best for $n=3$ and becomes worse with
decreasing $n$, both for low and high temperatures. A quantitative
measure for the agreement is given by the $\chi^2$-values for $n \in
\{1, 2, 3\}$, which are displayed in each of the panels in
Fig.~\ref{fig:gammam(T)}.  This conclusion is in accordance with
\pprI{}, where the $n=3$ case also agreed best
with the experimental data, although $\TK$ and $\gammam (T)$ for $n=3$
were significantly less accurate then.

\begin{figure*}
\includegraphics[height=16.5cm]{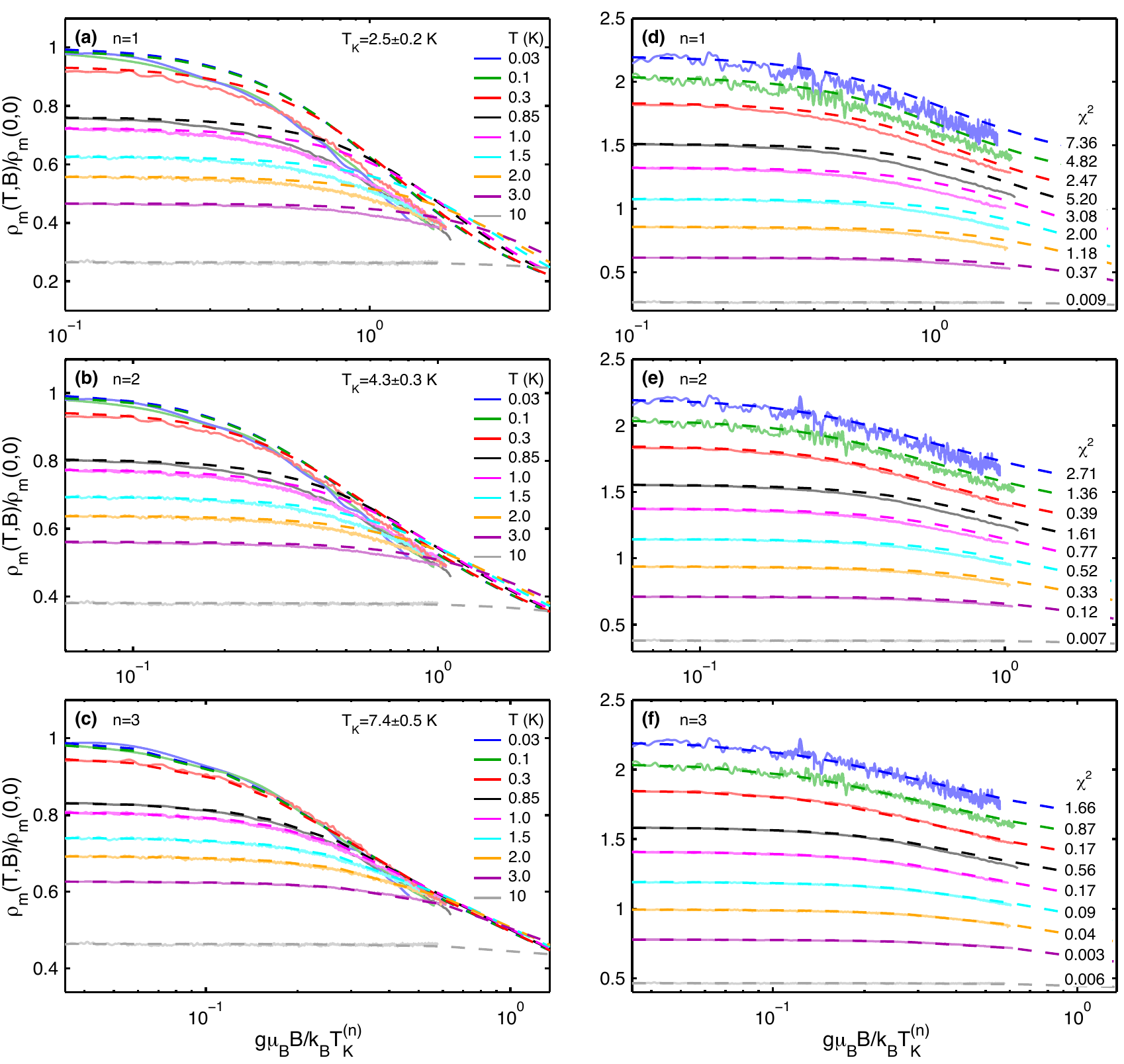}
\caption{(Color online) Experimental and theoretical results for $\rhom(T,B)$, shown
  using solid or dashed curves, respectively.  Left column: panels
  (a), (b) and (c) compare the experimental data for AgFe2 to
  NRG calculations for $n \in \{1, 2, 3\}$, respectively. Right
  column: panels (d), (e) and (f) show the same data as in the left
  column, except that for clarity the curves for successive
  temperatures are shifted vertically by 0.15 to avoid them from
  overlapping, thus enabling a better comparison between experiment
  and theory for each temperature.  $\TKrn$ and $\delta^{(n)}$ are
  already determined by the fitting procedure of
  Eq.~(\ref{eq:rhoNRG}), which allows a parameter-free theoretical
  prediction for $\rhom(T,B)$. The $\chi^2$-values indicated in
  panels (d-f) were calculated using a set of 1000 uniformly-spaced 
  field values in the range $B \in [0.07349,3.05000] {\rm T}$.
  The experimental data clearly show best
  agreement with theory for $n=3$, which supports the conclusion from
  the examination of $\gammam$. For $T=0.030$ K and $T=0.10$ K,
  the signal to noise ratio is much lower than for the other
  curves since the measurement current had to be reduced to stay in
  thermal equilibrium; therefore, in the left panels the experimental
  data for these two temperatures have been smoothed for better
  visibility.  For the largest temperature, $T=10$ K, the phonon
  contribution has been subtracted from the experimental data for
  comparison to theory. For the purpose of this subtraction, the
  phonon-contribution was assumed to be $B$-independent and taken to
  correspond to the difference of $\Delta\rho(T=10 K,
  B=0)/\Delta\rho(0,0)$ between experiment and theory (see
  Fig.~\ref{fig:rho(T)}).}
\label{fig:rho(T,B)}
\end{figure*}

Next we turn to the magnetoresistivity.  The above-mentioned
implementation of non-Abelian symmetries in our NRG code
\cite{Weichselbaum2012}, which drastically reduces computation time
and memory requirements, allows us to extend the analysis of
$\rhom(T)$ of \pprI{} to the whole two-dimensional parameter space of
$T$ and $B$. Since the fitting procedure of $\rhom(T,B=0)$ described
above leaves no further free parameters, this comparison is an
additional strong check of the validity of the $n=3$ model.  The
experimental data of $\rhom(T,B)$ for the sample AgFe2 are shown
together with the numerical data for $n \in \{1, 2, 3\}$ in
Fig.~\ref{fig:rho(T,B)}. [The values of $\rhom(T,B=0)$ differ for $n
\in \{1, 2, 3 \}$, due to the different $\delta^{(n)}$-values
determined from Eq.~(\ref{eq:rhoNRG}).] Again, the three-channel model
reproduces the measured results best.  Even though there are still
slight deviations between theory and experiment at high magnetic field
for the $n=3$ curves at 0.1~K and 0.85~K, which might originate from
very small temperature drifts, the overall agreement, combined with
that for $\gamma_m(T)$ (Fig.~\ref{fig:gammam(T)}) and $\rhom(T,0)$
(Fig.~\ref{fig:rho(T)}), is rather impressive. Thus, we conclude that
the $n=3$ model consistently reproduces all the transport data
discussed above.
  
\begin{figure}
\includegraphics[width=\linewidth]{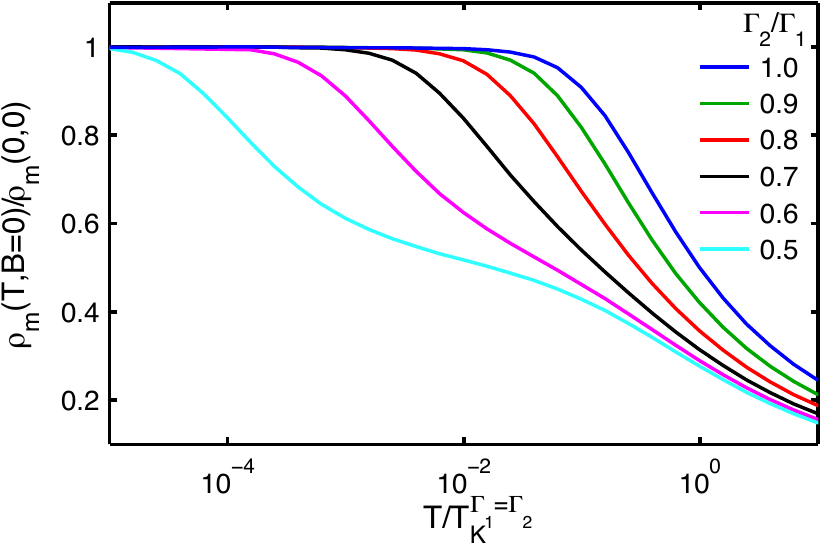}
\caption{(Color online) Temperature dependence of the resistivity for a 
  channel-anisotropic Kondo model with $S=1$,
  $n=2$, for several different choices of
  $\Gamma_2/\Gamma_1$.}
\label{fig:anisotropy}
\end{figure}

\subsection{Channel anisotropy}

To conclude this section, let us briefly discuss the possibility that
the true effective Kondo model for Fe in Au and Ag could include some channel
anisotropy. 

Channel anisotropy, if present at all, will be weak for the
present system due to a symmetry agrument. As mentioned in the
introduction, Fe acts as substitutional defect in Au or Ag; it hence
finds itself in an environment with cubic symmetry. This cubic
symmetry protects the equivalence of the three local $t_{2g}$ levels and of the
three bands involved in the effective low-energy Kondo model.  In
particular, this cubic symmetry offers a rather strong protection
against any splitting of the $t_{2g}$ levels. A significant spin-orbit coupling,
which could result in a spliting of the $t_{2g}$ levels, was ruled out by density
functional theory calculations in paper I.

With this in mind, let us nevertheless briefly discuss the
possible effects of channel anisotropy, that could arise if
some perturbation breaks the cubic symmetry.  In general, such a
perturbation could result in a small splitting in the $n$ impurity
$d$-levels that yield the spin $n/2$, or in slightly different band widths or
density of states for the $n$ conduction-band channels, or in slighlty 
different coupling strenghts between local and band states in each
channel.  All of these will will have similar effects on the
low-energy Kondo physics.

For concreteness, we consider here only the latter case, implemented
in our model by setting $t \to t_\alpha$ in Eq.~(1), leading to
channel-dependent level widths $\Gamma_\alpha = \pi \nu_0
t_\alpha$. For a spin-$n/2$, $n$-channel Kondo model, the presence
of channel anisotropy quickly leads to a multi-stage Kondo
effect,\cite{Nozieres1980,Affleck1992} characterized by $n$
different Kondo temperatures $\TKa$, in which channels of decreasing
$\Gamma_\alpha$ successively screen the bare spin $n/2$ first to
spin $(n-1)/2$, then to $(n-2)/2$, etc., down to 0. Since the
corresponding Kondo temperatures $\TKa$ depend exponentially on
$\Gamma_\alpha$, even a small amount of channel anisotropy changes
the shape of the resistivity curve $\rhom(T,B=0)$ drastically. In
particular, it spoils the purely logarithmic temperature dependence of
the resistivity for $T \simeq \TK$ that is characteristic of the
channel-isotropic Kondo effect: though each screening stage
separately produces a logarithmic contribution to the resistivity,
the sum of these contributions
no longer behaves purely logarithmically, as illustrated in
Fig.~\ref{fig:anisotropy} for $n=2$. Our experimental data, however,
do not show signatures of such multi-stage Kondo physics. This
implies that any channel anisotropy, if present, is
weak. Therefore the differences between the various
$\TKa$-values associated with the successive stages of screening
are, first, too small to be discernible in the data, and
second, not at all required for the interpretation of the
experimental data. We conclude that a fully channel-symmetric model
suffices.

\section{Conclusion}
\label{sec:conclusions}
We have considered iron impurities in gold and silver and compared
experimental data for the resistivity and decoherence rate to NRG
results for a fully screened $n$-channel, spin-$\frac{n}{2}$ Kondo
model. Compared to previous work on this subject \cite{Costi2009}, we
showed improved numerical data for both quantities at finite
temperature. In particular, we offered a detailed discussion 
of NRG convergence and truncation issues, using 
the discarded weight as a criterion
for reliably judging the quality of convergence.
Our most important new result
is the analysis of the resistivity at finite
magnetic field, where we  compare the numerical calculations with as yet
unpublished experimental data. In contrast to previous attempts to
explain the experimental results with models with less channels which
were inconsistent or yielded several different values
for the Kondo temperature, depending on which set of measurements was
used to extract $\TK$ \cite{Alzoubi2006}, we showed that all
examined quantities can be described consistently with a single 
value of $\TK$. The excellent agreement between experiment and 
theory for $n=3$ shows that both systems are well described by a spin-3/2 
three-channel Kondo model.

\section*{Acknowledgements}
We thank Norman Birge for helpful comments on the manuscript.
We gratefully acknowledge financial support
from ANR PNANO ‘‘QuSPIN’’ for L.S. and C.B.,
from the John von Neumann Institute for Computing
(J\"ulich) for T.C., from WE4819/1-1 for A.W., and
from SFB-TR12, SFB-631 and the Cluster of Excellence Nanosystems Initiative
Munich for J.v.D., M.H.\ and A.W.


\begin{thebibliography}{25}
\bibitem{analytics} The $n$-dependent differences in the shapes
    of the resistivity curve can be characterized in terms of
    analytic expressions for their asymptotic behavior at large and
    small ratios of $T/\TK^{(n)}$ and $g \mu_{\rm B} B / k_{\rm B}
    \TK^{(n)}$.  Such an analysis would go beyond the scope of the
    present paper and will be published elsewhere.
\end{thebibliography}
\end{document}